\documentclass[epj]{svjour}

\usepackage{graphicx}
\usepackage{amsmath}
\usepackage{bm}                      % bold math
\usepackage{xcolor}                  %
\def\bea{\begin{eqnarray}} \def\eea{\end{eqnarray}}
\def\be{\begin{equation}} \def\ee{\end{equation}}
\def\bal{\begin{align}} \def\eal{\end{align}}
\def\bse{\begin{subequations}} \def\ese{\end{subequations}}

\def\anss{Ans\"atze} % Plural of 'Ansatz' !

\def\al{\alpha}
\def\eps{\varepsilon}
\def\ms{M_\odot}
\def\bds{B_\text{DS}}

\begin{document}

\title{
Strange quark matter and quark stars with the Dyson-Schwinger quark model}

%\begin{CJK*}{GB}{gbsn} % Use default fonts from CJK
%\begin{CJK*}{GB}{SongMT}
%\begin{CJK*}{GB}{gbsn}

\author{
H. Chen \inst{1}
\and
J.-B. Wei \inst{1}
\and
H.-J. Schulze \inst{2}}

\institute{
School of Mathematics and Physics, China University of Geosciences,
Lumo Road 388, 430074 Wuhan, China
\and
INFN Sezione di Catania, Dipartimento di Fisica,
Universit\'a di Catania, Via Santa Sofia 64, 95123 Catania, Italy}

%\email{huanchen@cug.edu.cn}

\date{Received: date / Revised version: date}

\abstract{
We calculate the equation of state of strange quark matter
and the interior structure of strange quark stars
in a Dyson-Schwinger quark model
within rainbow or Ball-Chiu vertex approximation.
We emphasize constraints on the parameter space of the model
due to stability conditions of ordinary nuclear matter.
Respecting these constraints,
we find that the maximum mass of strange quark stars is about 1.9 solar masses,
and typical radii are 9--11 km.
We obtain an energy release
as large as $3.6 \times 10^{53}\,\text{erg}$
from conversion of neutron stars into strange quark stars.
}

% PACS numbers are not required since they are no
% longer published in EPJ journals !
%\PACS{
% 26.60.Kp,  % Equations of state of neutron star matter
% 12.39.-x,  % Phenomenological quark models
% 97.60.Jd,  % Neutron stars
% 12.39.Ba,  % Bag model
% 26.60.-c,  % Nuclear matter aspects of neutron stars
% 26.60.+c,  % Nuclear aspects of neutron stars
% 21.65.+f,  % Nuclear matter
% 24.10.Cn,  % Many-body theory
% 26.50.+x,  % Nuclear physics aspects of supernovae
% 26.60.Dd,  % Neutron star core
% 26.60.Gj,  % Neutron star crust
% 21.65.Mn,  % Equations of state of nuclear matter
% 21.65.Cd,  % Asymmetric matter, neutron matter
% 13.75.Cs,  % Nucleon-nucleon interactions
% 13.75.Ev   % Hyperon-nucleon interactions
% 26.50.+x,  % Nucl. physics aspects of (super)novae and explosive environments
% 97.10.Cv,  % Stellar structure, interiors, evolution, nucleosynthesis, ages
% 97.60.Gb,  % Pulsars
% 25.75.Nq,  % Quark deconfinement, QGP production, phase transitions in RHIC
% 12.38.Mh,  % Quark-gluon plasma in quantum chromodynamics
%Supernovae, 97.60.Bw
%Protostars, 97.21.+a}

\maketitle

%===============================================================================
\section{Introduction}

The hypothesis of stable strange quark matter (SQM)
\cite{Bodmer71,Witten84,Farhi84}
and strange quark stars (SQSs)
\cite{Ivanenko65,Itoh70,Terazawa79}
has been attracting interest since some time.
Originally it was demonstrated that in a wide region of the MIT bag model
parameter space, SQM, but not nuclear matter,
could form the ground state of baryonic matter.
Since then the SQM hypothesis has been addressed in numerous articles,
see \cite{gle,Weber05,Haensel:Book} for an overview.

The most probable way to observe SQM is in compact stars.
Compact stars built entirely of quark matter (QM) were studied
in different levels of sophistication,
starting with an equation of state (EOS) of a free degenerate Fermi gas
of $u,d,s$ quarks with equal masses.
Then the structure of SQSs was examined in detail in MIT-bag-type models,
taking into account the strange quark mass,
the lowest-order correction from the QCD interaction,
etc.~\cite{Haensel86a,Alcock86}.
More work on the EOS of SQM,
the formation of SQSs,
their neutrino emission, rotation, superfluidity, pulsations,
electromagnetic radiation, and cooling were also done,
see the reviews \cite{gle,Weber05,Bombaci01,Weber14}.
QM in the interior of massive neutron stars (NSs)
is one of the current main issues in the physics of compact stars,
due to the recent observations of two NSs of about two solar masses,
PSR~J1614-2230 ($M/\ms=1.93\pm0.02$) \cite{heavy,Fonseca16}
and
PSR~J0348+0432 ($M/\ms=2.01\pm0.04$) \cite{heavy2}.

The EOS of QM is crucial for the study of SQM and SQSs.
Many works have been done to go beyond the MIT bag model,
e.g., using perturbative QCD
\cite{Baluni78,Fraga01,Kurkela10,Fraga14,Xu15},
the density-dependent-quark-mass model
\cite{Fowler81,Chak91,Benvenuto95,Liang10,Torres13},
the Nambu-Jona-Lasino model
\cite{Buballa96,Buballa99,Buballa05,Schertler99,Klahn,Klahn15},
the chiral quark meson model
\cite{Zacchi15},
or the quasi-particle model
\cite{Peshier00,Tian12,Zhao15}.
However,
the EOS remains poorly known due to the nonperturbative character of QCD.
The Dyson-Schwinger equations (DSE) provide a continuum
approach to QCD that can simultaneously address both confinement and
dynamical chiral symmetry breaking
\cite{Roberts1994dr,Alkofer2000wg}.
They have been applied with success to hadron physics in vacuum
\cite{Roberts2007jh,Fischer09a,Chang2009zb,Chang13,Eichmann2009,Eichmann10,Eichmann16}
and to QCD at nonzero temperature and chemical potential
\cite{Roberts2000aa,Maas2013,Fischer09b,Fischer09c,Fischer14,Fischer2013,Fischer2014b,%
Eichmann2016,Qin2010nq,Gao2014,Gao16,Nickel06a,Nickel06b,Muller2013a,Muller16,%
Chen2008zr,Chen2011,Chen12,Chen15a,Chen15b,Zhao2015}.
Both MIT and NJL model have been recognized
as limiting cases of the DSM \cite{Klahn,Klahn15}.

In this paper,
we use a Dyson-Schwinger model (DSM) for QM based on our
previous work \cite{Chen2011,Chen12,Chen15a,Chen15b},
in which the hadron-quark phase transition in compact stars
and the structure of hybrid stars were investigated,
in combination with a nuclear matter EOS within the Brueckner-Hartree-Fock
(BHF) many-body approach \cite{bbb,zhou,mmy,zhli}.
However, there are still free parameters due to uncertainties
of the effective interaction and vacuum pressure in our model,
and we will scan its parameter space to
investigate the possibility of SQM and SQSs.

The paper is organized as follows.
In section \ref{s:qm} we briefly discuss the DSM for QM
and the parameters in our model.
In section \ref{s:sqm} we analyze the allowed parameter space for stable SQM,
and the corresponding EOS.
In section \ref{s:sqs} we present results on the structure of SQSs,
as well as the energy release from conversion of NSs into SQSs.
Section \ref{s:end} contains our conclusions.

%===============================================================================
\section{Formalism: Quark matter with the Dyson-Schwinger model}
\label{s:qm}

%\subsection{Quark Matter with the Dyson-Schwinger model}

\begin{figure}[t]%..............................................................
%\vspace{-12mm}
\centerline{\includegraphics[scale=0.5]{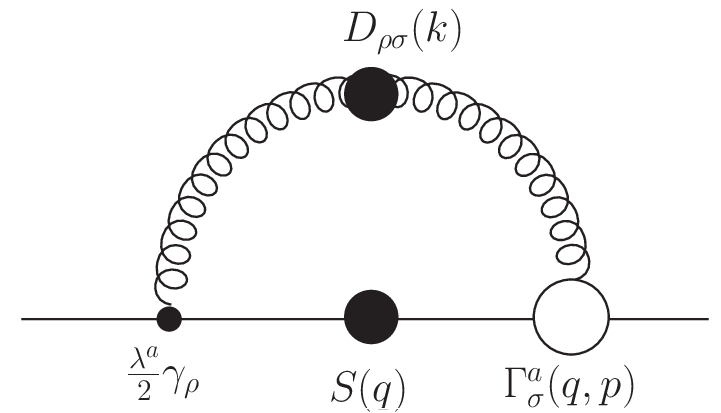}}
%\vspace{-2mm}
\caption{(Color online)
Diagrammatic representation of the quark self-energy,
Eq.~(\ref{e:gensigma}),
in the DSM.}
\label{f:dse}
\end{figure}%...................................................................

For cold dense QM,
we adopt a model based on the DSE of the quark propagator,
described in detail in our previous papers \cite{Chen2011,Chen12,Chen15a,Chen15b}.
In the following, we only give a brief introduction to the model.
We start from the gap equation for the quark propagator $S(p;\mu)$
at finite chemical potential
$\mu\equiv\mu_q=\mu_B/3$,
depicted in Fig.~\ref{f:dse},
\be
 \Sigma(p;\mu) =
 \int\! \frac{\text{d}^4p}{(2\pi)^4} \,
 %g^2(\mu)
 S(q;\mu)
 \frac{\lambda^a}{2} \gamma_\rho
 D_{\rho\sigma}(k;\mu)
%\non\\&&\times
 %\frac{\lambda^a}{2}
 \Gamma^a_\sigma(q,p;\mu) \:,
\label{e:gensigma}
\ee
where $\lambda^a$ are the Gell-Mann matrices, $k=p-q$,
%$g(\mu)$ is the coupling strength,
$D_{\rho\sigma}(k;\mu)$ is the dressed gluon propagator,
and $\Gamma^a_\sigma(q,p;\mu)$ the dressed quark-gluon vertex
at finite chemical potential.
To solve the equation,
one requires an Ansatz for both $D_{\rho\sigma}$ and $\Gamma^a_\sigma$.

In our model,
the combined Ansatz for $D_{\rho\sigma}$ and $\Gamma^a_\sigma$
is parameterized as
\be
 %Z_1 g^2
 D_{\rho \sigma}(k) \Gamma_\sigma^a(q,p) =
%\non\\&&\hskip15mm
 {\cal G}(k) \, D_{\rho\sigma}^{\rm free}(k)
 \frac{\lambda^a}{2}\Gamma_\sigma(q,p) \:,
\label{e:KernelAnsatz}
\ee
wherein
$D_{\rho\sigma}^{\rm free}(k) =
\big(\delta_{\rho\sigma}-\frac{k_\rho k_\sigma}{k^2}\big)
\frac{1}{k^2}$
is the Landau-gauge free gluon propagator
and $\Gamma_\sigma(q,p)$ represents the tensor structure
of the quark-gluon vertex Ansatz,
while other dressing effects of both the vertex and the gluon propagator
are assumed to depend only on the gluon momentum $k$ and
are included in a model effective interaction ${\cal G}(k)$.
Herein we neglect many effects,
such as the violation of Lorentz covariance of the gluon propagator
at finite chemical potential,
the possibility of color-superconductivity
\cite{Nickel06a,Nickel06b,Muller2013a},
etc.

For $\Gamma_\sigma$,
three forms were investigated in our previous work \cite{Chen15a,Chen15b}:
(1) the bare vertex, also called rainbow (RB) approximation.
(2) the Ball-Chiu (BC) vertex,
which satisfies the Ward-Takahashi identity of QED
and is free of kinetic singularity.
The form of the BC vertex in vacuum was given in \cite{ball-chiu},
and was extended to finite chemical potential in \cite{Chen2008zr}.
(3) the 1BC vertex, which includes only part of the BC vertex,
but is numerically quite similar to the RB approximation.
Therefore, in this work we will only present results with the
RB and BC approximations.

For the effective interaction,
we employ an infrared-dominant interaction
modified by the quark chemical potential \cite{Chen2011,Chen12,Jiang2013}
\be
 \frac{{\cal G}(k)}{k^2}  =
 4\pi^2 d \frac{k^2}{\omega^6} e^{-\frac{k^2+\al\mu^2}{\omega^2}}
  \:.
\label{gaussiangluonmu}
\ee
The parameters $\omega$ and $d$ can be obtained by fitting meson properties
and chiral condensate in vacuum \cite{Chang2009zb,Chang13,Alkofer2002bp},
and we use $\omega=0.5\;\text{GeV}$,
$d=1\;\text{GeV}^2$ (with RB),
$d=0.5\;\text{GeV}^2$ (with BC).
We choose the quark masses
$m_{u,d}=0$ and $m_s=115\;\text{MeV}$.
(We discuss later a possible variation of $m_s$).
The phenomenological parameter $\al$ is of particular importance in our work,
since it represents a reduction rate of the
effective interaction with increasing chemical potential.
However, it cannot yet be fixed independently.
Obviously,
$\al=\infty$ corresponds to a noninteracting system at finite chemical potential,
i.e., a simple version of the MIT bag model,
but in previous and present work
we investigate the full parameter space $0<\al<\infty$.
With varying value of $\al$,
we can investigate the role of the interaction strength and
confront it with different mechanism from other phenomenological models.
For example, comparing with the quasi-particle model,
we found that the effects of the interaction on light quarks
are not in the form of an effective mass,
but a modification of the vector part of the quark propagator \cite{Chen2011,Chen12}.

\begin{figure}[t]%..............................................................
\vspace{-6mm}
\centerline{\includegraphics[scale=0.49]{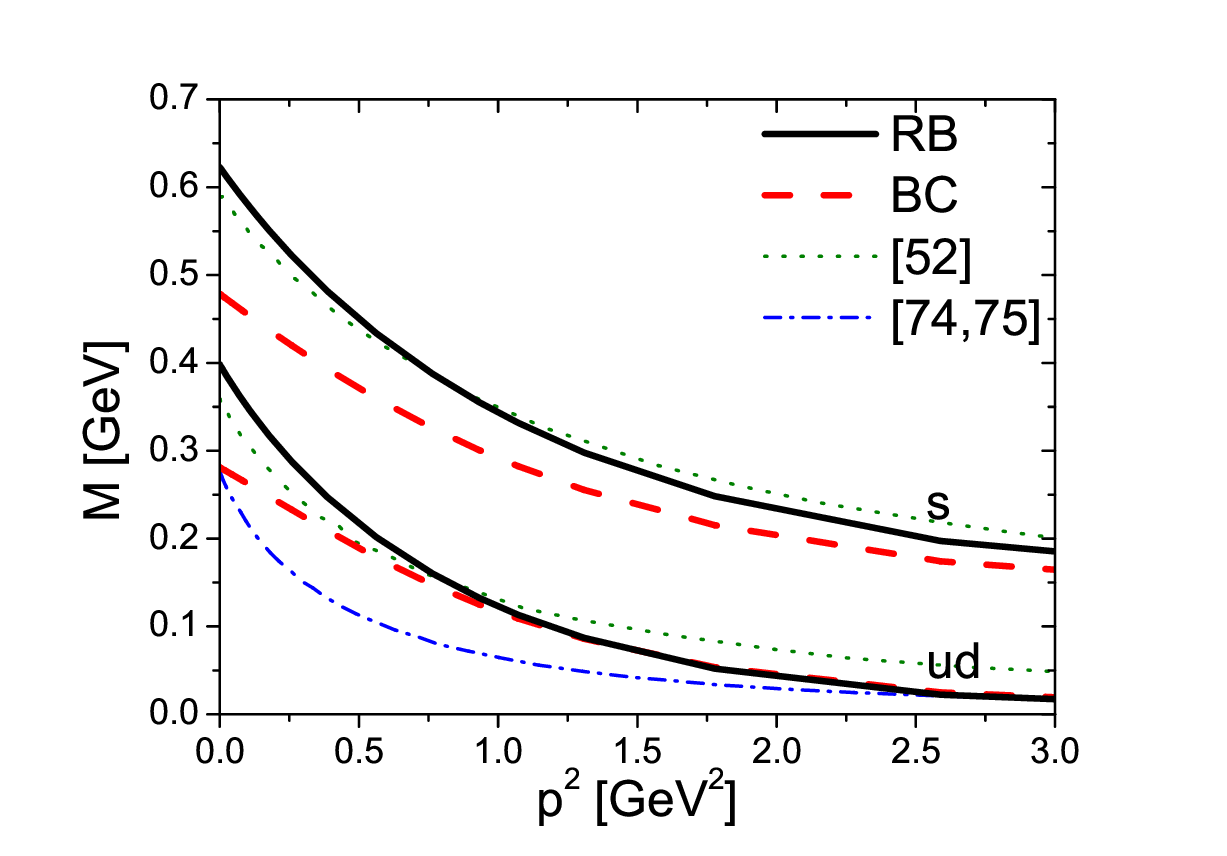}}
\vspace{1mm}
\caption{(Color online)
Quark propagator vacuum mass function $M(p^2)$,
Eq.~(\ref{e:sinvp}),
for $u,d$ and $s$ quarks
and for different vertex \anss\ RB and BC,
in comparison with results of
Refs.~\cite{Fischer2014b} and \cite{Bhagwat2003,Roberts08}.
}
\label{f:mpp}
\end{figure}%...................................................................

The general structure of the quark propagator at finite chemical potential is
\be
 S(p;\mu)^{-1} =
 i {\bm \gamma}{\bm p}A(p^2\!\!,p_4) + B(p^2\!\!,p_4)
%\non\\&&
 + i \gamma_4(p_4+i\mu)C(p^2\!\!,p_4) \:,
\label{e:sinvp}
\ee
where
$A,B,C$ are complex scalar functions and $C= A$ at $\mu=0$.
The mass function $M \equiv B/A$ is of particular interest for physical
interpretation (asymptotic freedom and dynamical mass generation),
and we show the quantity $M(p^2)$ of the Nambu (confined) solution at $\mu=0$
in Fig.~\ref{f:mpp},
in comparison with the corresponding results of
Refs.~\cite{Fischer2014b,Bhagwat2003,Roberts08},
which are qualitatively similar.
The latter reference also contains a confrontation with lattice QCD results.
%The comparison shows that the mass functions in our model decrease more quickly
%in the ultraviolet (UV) region,
%due to neglecting the UV interaction in Eq.~(\ref{gaussiangluonmu}).
%In the infrared region,
%results with RB approximation are a little higher than in \cite{Fischer2014b},
%which could be expected to complement the missing UV part.
Results with BC vertex feature reduced interaction effects
and lower mass functions.
%Results for chiral quarks from Ref.~\cite{Bhagwat2003} are much lower than others.

\begin{figure}[t]%..............................................................
\vspace{-12mm}
\centerline{\includegraphics[scale=0.42]{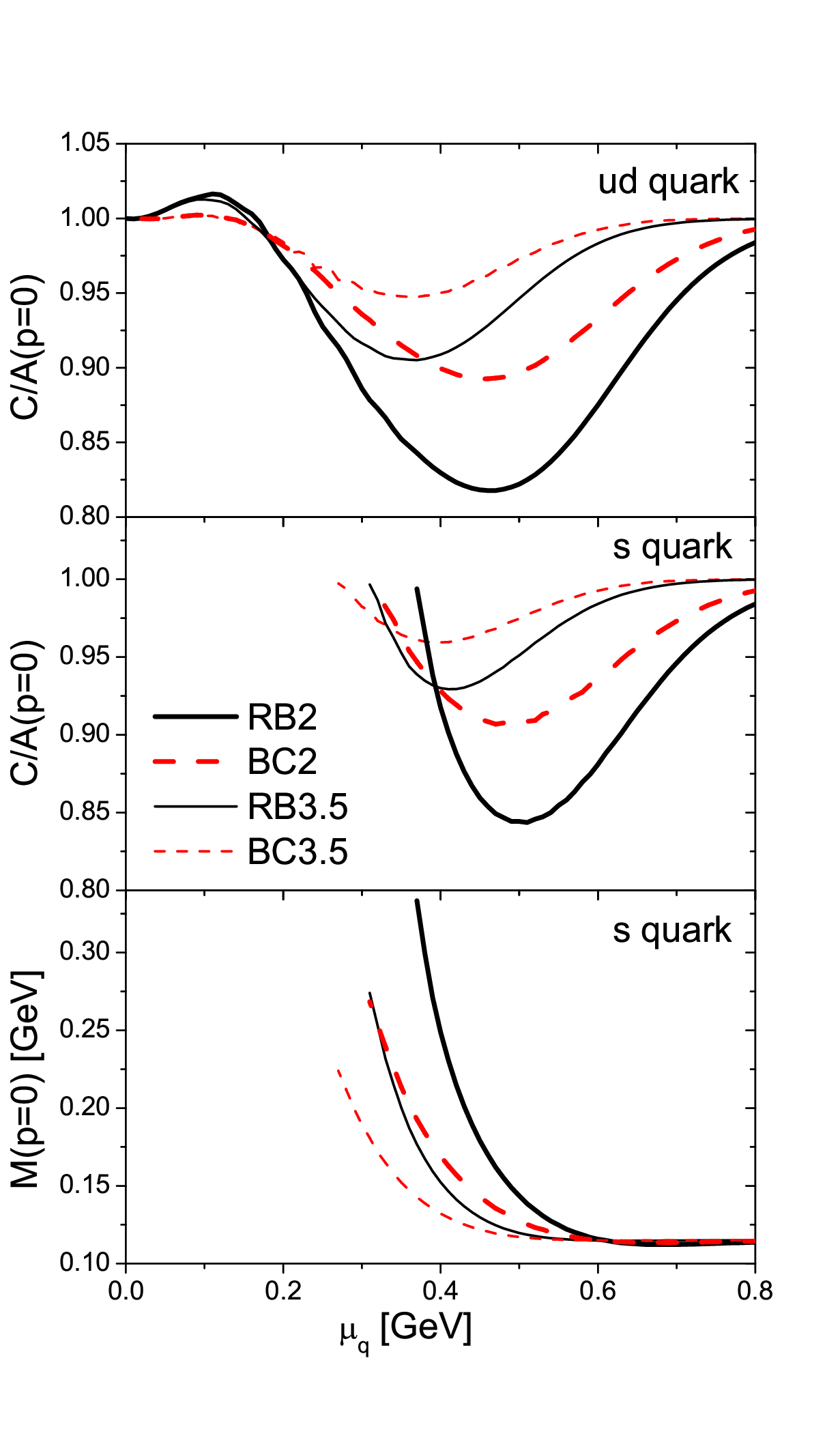}}
\vspace{-6mm}
\caption{(Color online)
Wigner solutions of the
quark propagator components $C/A$ and $M$ at $p=0$
for $u,d$ and $s$ quarks
at finite chemical potential
for different vertex \anss\ RB and BC
and for $\al=2, 3.5$,
using the notation RB$\al$, BC$\al$.}
\label{f:mmu}
\end{figure}%...................................................................

At finite chemical potential the discussion of the propagator becomes
more complex,
due to the additional dependence on $p_4$ and the presence of the function $C$.
In this work we just illustrate the behavior in Fig.~\ref{f:mmu}
by showing the dependence on chemical potential
of the components $C/A$ and $M=B/A$ at $p=0$
for the Wigner (deconfined) solution.
%which are real due to the condition $F(p^2,p_4)=F^*(p^2,-p_4)$.
As is shown in \cite{Chang2007,Wang2012},
the Wigner solution exists always for massless quarks (upper panel),
but for massive strange quarks (lower panels)
only at high enough chemical potential,
as is seen in the figure.
At low chemical potential only the Nambu solution exists,
which is characterized by $C\equiv A$
and quite large effective mass \cite{Chen2008zr}.
For massless $u,d$ quarks one has $B=M=0$, and
one can see that $C/A(p=0)<1$,
apart from a small region at low chemical potential.
This causes the effective Fermi momentum,
and consequently the quark number density,
to be lower than for free quarks at the same chemical potential
\cite{Chen2008zr,Chen2011,Chen12}.
For the strange quarks, both $C/A<1$ and $M>m_s$
lead to the same effects on the quark number density.
Choosing the BC Ansatz or increasing the parameter $\al$
leads to weaker interaction effects,
as can be seen clearly.

All the relevant thermodynamical quantities of cold QM can be computed
from the quark propagator at finite chemical potential,
except a boundary value of the pressure $P$,
which is represented by a phenomenological bag constant $\bds$
that we consider another model parameter,
\be
 P(\mu_u,\mu_d,\mu_s) = - \bds + \sum_{q=u,d,s}
 \int_{\mu_q^0}^{\mu_q}\! {\rm d}\mu \,n_q(\mu) \:,
\ee
where the density distributions $n_q$ are obtained from the quark propagator
\cite{Chen2008zr,Chen2011,Chen12,Klahn2009mb},
\bea
 n_q(\mu) &=& 6 \int\frac{d^3 p}{(2\pi)^3} \, f_q(|\bm{p}|;\mu) \:,
\label{nqmu}
\\
 f_q(|\bm{p}|;\mu) &=&
 \frac{1}{4\pi} \int_{-\infty}^\infty \! dp_4 \,
 {\rm tr}_{\rm D}\big[-\gamma_4 S_q(p;\mu)\big] \:,
\label{nqmuf1}
\eea
and the trace is over spinor indices only.
Detailed results for the density distributions were shown and discussed in
Refs.~\cite{Chen2011,Chen12,Chen15a,Chen15b}.

In principle, $\bds$ can be obtained from the pressure difference between the
deconfined (Wigner) phase and the confined (Nambu) phase in vacuum, i.e.,
\be
 \bds = \sum_{q=u,d,s} \left[ P_q^{(c)}(\mu_q=0) - P_q^{(d)}(\mu_q^0) \right]
\:.
\ee
In the framework of DSEs,
one usually uses the `steepest-descent' approximation \cite{haymaker1990vm}
to calculate the pressure
\be
 P[S] =  {\rm TrLn}\left[S^{-1}\right] -
 \frac{1}{2}{\rm Tr}\left[\Sigma\,S\right] \:,
\label{e:pSigma}
\ee
and in this way we obtain at $\mu=0$ a pressure difference
for massless quarks within RB approximation \cite{Chen2008zr}
\be
 P^{(c)}_{u,d}(\mu=0) - P^{(d)}_{u,d}(\mu=0) \approx 45\;{\rm MeV\,fm^{-3}} \:.
\label{e:calb}
\ee
Interpreting the Nambu phase as the real vacuum with
$P^{(c)}_{u,d}(\mu=0)=0$,
we then obtain the pressure of the Wigner phase for light quarks in vacuum
$P^{(d)}_{u,d}(\mu_0=0) \approx -45\;{\rm MeV\,fm^{-3}}$
and the effective bag constant from contributions of $u$ and $d$ quarks as
$\bds^{n_f=2} \approx 90\;{\rm MeV\,fm^{-3}}$.
This value was used in our previous work.
However, such an approximation is only consistent with the RB approximation
and a static gluon propagator.
There is no Wigner solution for strange quarks at $\mu=0$
\cite{Chang2007,Jiang12}.
With the introduction of the parameter $\al$ only in the Wigner phase,
the steepest-descent approximation is not consistent with the gap
equation at finite chemical potential.
So we cannot obtain $P_s(\mu_{s,0})$ and its contribution to $\bds$ is unclear.
Furthermore, the steepest-descent approximation is not consistent
with the BC vertex either.
Therefore, in this paper we allow a free variation of $\bds$,
but expecting it to be of the same order as $90\;{\rm MeV\,fm^{-3}}$.
In the following, $\bds$ is always given in units of
$\text{MeV}\,\text{fm}^{-3}$ in the text and figures.

\begin{figure}[t]%..............................................................
\vspace{-12mm}
\centerline{\includegraphics[scale=0.51]{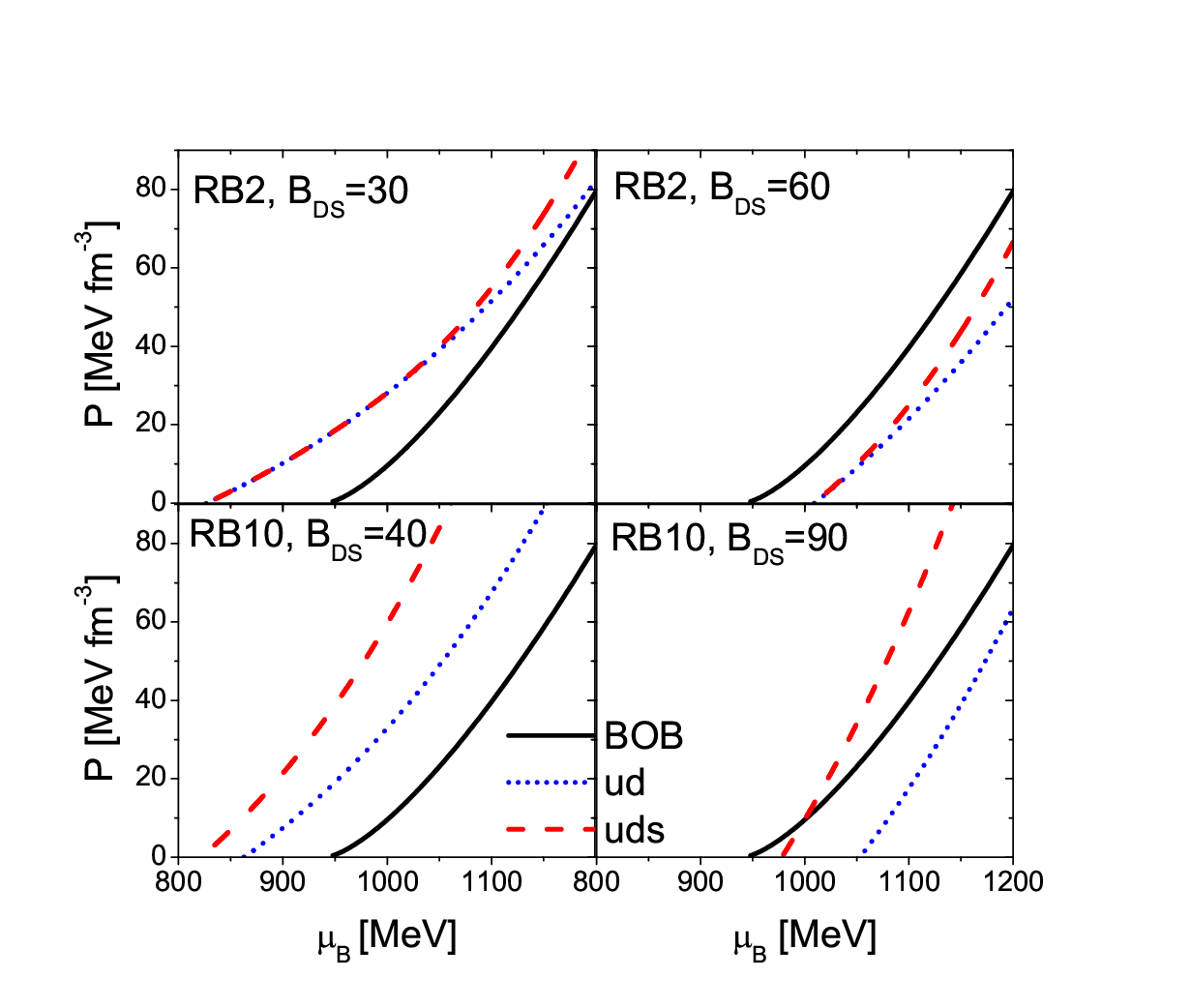}}
\vspace{-3mm}
\caption{(Color online)
Pressure vs.~baryon chemical potential of dense NS matter
for different models
[nuclear matter (BOB), 2-flavor (ud) and 3-flavor (uds) quark matter]
and parameters
[notation RB$\al$ indicates RB approximation with density parameter $\al$;
$\bds$ the value of the bag constant in $\text{MeV}\,\text{fm}^{-3}$].}
\label{f:1}
\end{figure}%...................................................................

%===============================================================================
\section{Results and discussion}

\subsection{Stability of strange quark matter}
\label{s:sqm}

Since SQM would probably appear in the form of SQSs \cite{Witten84},
we investigate in the following NS matter, i.e.,
cold, neutrino-free, charge-neutral,
and beta-stable SQM \cite{Chen2011,Chen12},
characterized by two degrees of freedom,
the baryon and charge chemical potentials $\mu_B$ and $\mu_Q$.
The corresponding equations are
\be
 \mu_i = b_i \mu_B + q_i \mu_Q \:,\quad
 \sum_i q_i \rho_i = 0 \:,
\ee
$b_i$ and $q_i$ denoting baryon number and charge of
the particle species $i=n,p,e,\mu$ in the nuclear phase
and $i=u,d,s,e,\mu$ in the quark phase, respectively.

In Fig.~\ref{f:1} we first illustrate the corresponding EOS $P(\mu_B)$
in our DSM with the RB approximation
(`uds', dashed red curves)
with various typical values of the parameters $\al$ and $\bds$
(different panels RB$\al$-$\bds$),
in comparison with the EOS of two-flavor QM (2QM)
(`ud', dotted blue curves),
as well as that of hadronic nuclear matter from the Brueckner-Bethe-Goldstone
theory with the Bonn-B potential \cite{zhli}
(`BOB', solid black curve).
At given chemical potential $\mu_B$,
the physically realized phase is the one with the highest pressure $P$.

With a parameter $\al=2$ (upper panels), i.e.,
a moderate reduction rate of the interaction strength,
strange quarks appear only at very large chemical potential
due to their large dynamical mass.
In such cases,
stable matter with zero pressure can only be 2QM
(upper left panel)
for a small effective bag constant $\bds$,
or nuclear matter (upper right panel)
for a large $\bds$.
An increasing value of $\bds$ reduces the stability of QM.

With a large reduction rate of the interaction strength,
$\al=10$ (lower panels),
the dynamical mass of strange quarks decreases quickly
and they can appear at small chemical potential.
In such cases, if $\bds$ is too small
(lower left panel),
nuclear matter at $P=0$ would be unstable against 2QM,
which is inconsistent with physical reality.
When $\bds$ is too large
(lower right panel),
at $P=0$ nuclear matter would be stable against 2QM,
but also against SQM.
In all the above cases, the hypothesis of SQM would not be valid.

\begin{figure}[t]%..............................................................
%\vspace{11mm}
\includegraphics[bb=1 100 800 1070,scale=0.3]{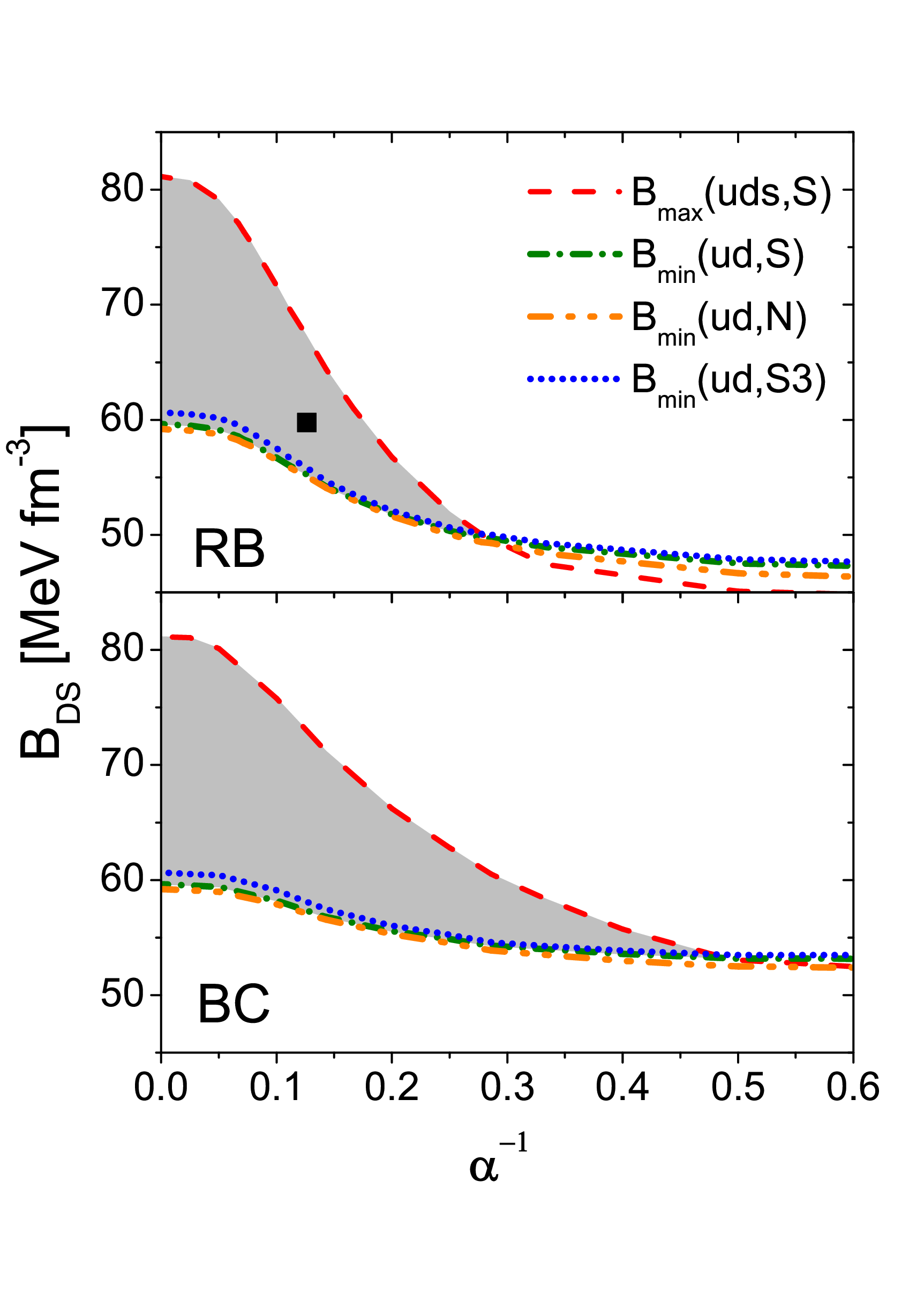}
%\vspace{-10mm}
\caption{(Color online)
The allowed parameter space of $\al$ and $\bds$ for the SQM hypothesis
in the DSM with RB (upper panel) or BC (lower panel) approximation,
respecting the nuclear matter stability conditions.
The various boundary curves,
$B_\text{min}$ indicating the lower limits due to the stability
of ordinary nuclear matter,
and $B_\text{max}$ the upper limit for the stability of SQM,
are further discussed in the text.
The marker refers to the choice of parameters for Fig.~\ref{f:pmusqm}.
}
\label{f:ab}
\end{figure}%...................................................................

However, the parameters $\al$ and $\bds$ cannot be chosen arbitrarily,
but are subject to severe constraints due to the fact that
ordinary stable baryonic matter in our world is non-strange nuclear matter.
We therefore have at least two quantitative constraints on the 2QM,
\bse
\bea
 \mu_B^\text{ud,S}(P=0) &>& 924\,\text{MeV} \:,
\\
 \mu_B^\text{ud,N}(P=0) &>& 939.4\,\text{MeV} \:,
\eea
\label{Eq:sn}%
\ese
obtained from the stability of symmetric nuclear matter (S)
and neutral nuclear (neutron) matter (N) \cite{Haensel:Book}.
These two conditions enforce lower bounds on the parameter $\bds$
for each $\al$.
On the other hand, for the hypothesis of SQM to be valid,
SQM should be stable with respect to ordinary nuclear matter in an iron nucleus,
\be
 \mu_B^\text{uds,S}(P=0) < 930.4\,\text{MeV} \:,
\label{Eq:sqm}
\ee
which determines an upper bound on the parameter $\bds$ for each $\al$.

These constraints on the parameters $\bds$ and $\al$
are visualized in Fig.~\ref{f:ab},
where the shaded area contains the values ($\al,\bds$) that produce SQM
according to Eq.~(\ref{Eq:sqm})
(upper dashed red boundary curve),
while respecting the stability conditions Eqs.~(\ref{Eq:sn}a,b)
(lower dash-dotted green and dash-dot-dotted orange boundary curves).
One can see that these two lower boundary lines are quite close to each other.
In fact, even if we tighten the stability constraint
and demand symmetric nuclear matter to remain in the nucleonic (BHF BOB) phase
up to 3 times saturation density (dotted blue curve)
the lower boundary does not change very much.

%Comparing the RB (top panel) and BC (bottom panel) approximations,
%the allowed parameter space is a little larger in the latter case:
One notes that the figure establishes
the absolute parameter bounds for possible SQM
$\al>3.6$, $50<\bds<81$ (RB, top panel)
and
$\al>2.1$, $53<\bds<81$ (BC, bottom panel).     %
In the MIT limit ($\al^{-1}=0$) the bounds are
$59<\bds<81$,                                   %  B^1/4= 146-158
and assuming massless strange quarks,
one obtains $59<\bds<92$ \cite{Haensel:Book}.   %  B^1/4= 146-163
From the plot one may conclude that an increasing interaction strength
in the DSM effectively reduces the stability of QM,
such that only small values of $\bds$ are permitted for increasing $\al^{-1}$.
This is also consistent with the larger allowed parameter space
with BC compared to RB approximation,
since the former features a weaker interaction
at finite chemical potential \cite{Chen15a,Chen15b}.

In particular, the DSM without in-medium dampening of the interaction
($\al=0$) does not allow SQM,
as it does neither provide the possibility of hybrid NSs \cite{Chen2011,Chen12}.
This feature was also exposed in Ref.~\cite{Chen2011,Chen12}
by demonstrating that the density-dependent bag parameter
\be
 B(\rho) \equiv \eps(\rho) - \eps_\text{free}(\rho)
\ee
is a rapidly rising function of density,
indicating the repulsive nature of the in-medium quark-quark interaction.
This is in clear contrast with the MIT model, for example,
where $B(\rho)$ is a constant by definition.

The lower boundaries in Fig.~\ref{f:ab} represent universal
stability conditions of ordinary nuclear matter
that have to be respected by any quark model,
whether modeling SQSs or ordinary (hybrid) quark NSs.
In the latter case,
there is a different upper limit $B_\text{max}$,
beyond which a transition to QM does not occur any more
even for the heaviest NSs, i.e.,
hybrid NSs cannot be formed at all with the given quark model.
This upper limit depends obviously on the chosen hadronic EOS
and is much larger than the upper limit for the SQM phase,
because the relevant phase transition might occur
at very large pressure.
%and we show in the figure the upper limit obtained with the BOB EOS
In fact, in the case of the BOB EOS for the hadronic phase,
this upper limit is beyond any reasonable range of the bag constant.
%One thus recognizes the regions of parameter space
%where SQSs or hybrid NSs may exist, and where no possibility is allowed...
For the same reason,
the lower limit on the parameter $\al$
is much lower for ordinary hybrid stars than for SQSs, e.g.,
in Ref.~\cite{Chen2011,Chen12} an $\al_\text{min}\approx0.5$ was considered.

\begin{figure}[t]%..............................................................
\vspace{-5mm}
\includegraphics[scale=0.16]{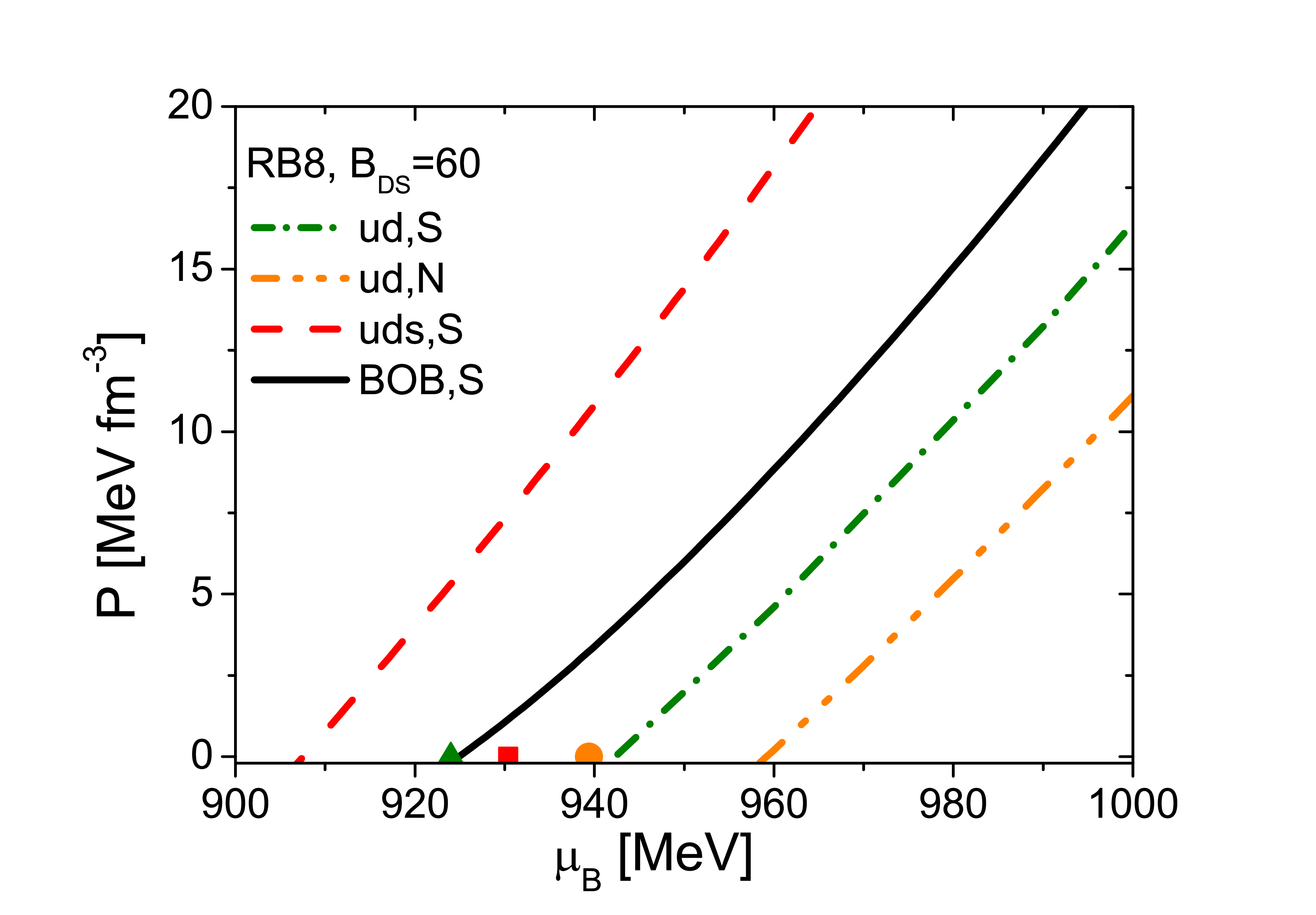}
\vspace{-1mm}
\caption{(Color online)
Pressure vs.~baryon chemical potential of various EOSs.
The stability conditions Eqs.~(\ref{Eq:sn}) and (\ref{Eq:sqm})
are symbolized by the markers.}
\label{f:pmusqm}
\end{figure}

Having determined the possible parameter values $\al$ and $\bds$,
we illustrate a typical result in Fig.~\ref{f:pmusqm},
which shows the low-density EOS of
symmetric nuclear matter (BOB,S),
symmetric SQM (uds,S), and
symmetric (ud,S) and neutral (ud,N) 2QM.
The RB approximation and the parameter values
$\al=8$, $\bds=60$
(see the black square marker in Fig.~\ref{f:ab})
are used.

%...................................................................

%===============================================================================
\subsection{Structure of strange quark stars}
\label{s:sqs}

\begin{figure}[t]%..............................................................
\vspace{-1mm}
\centerline{\includegraphics[scale=0.68]{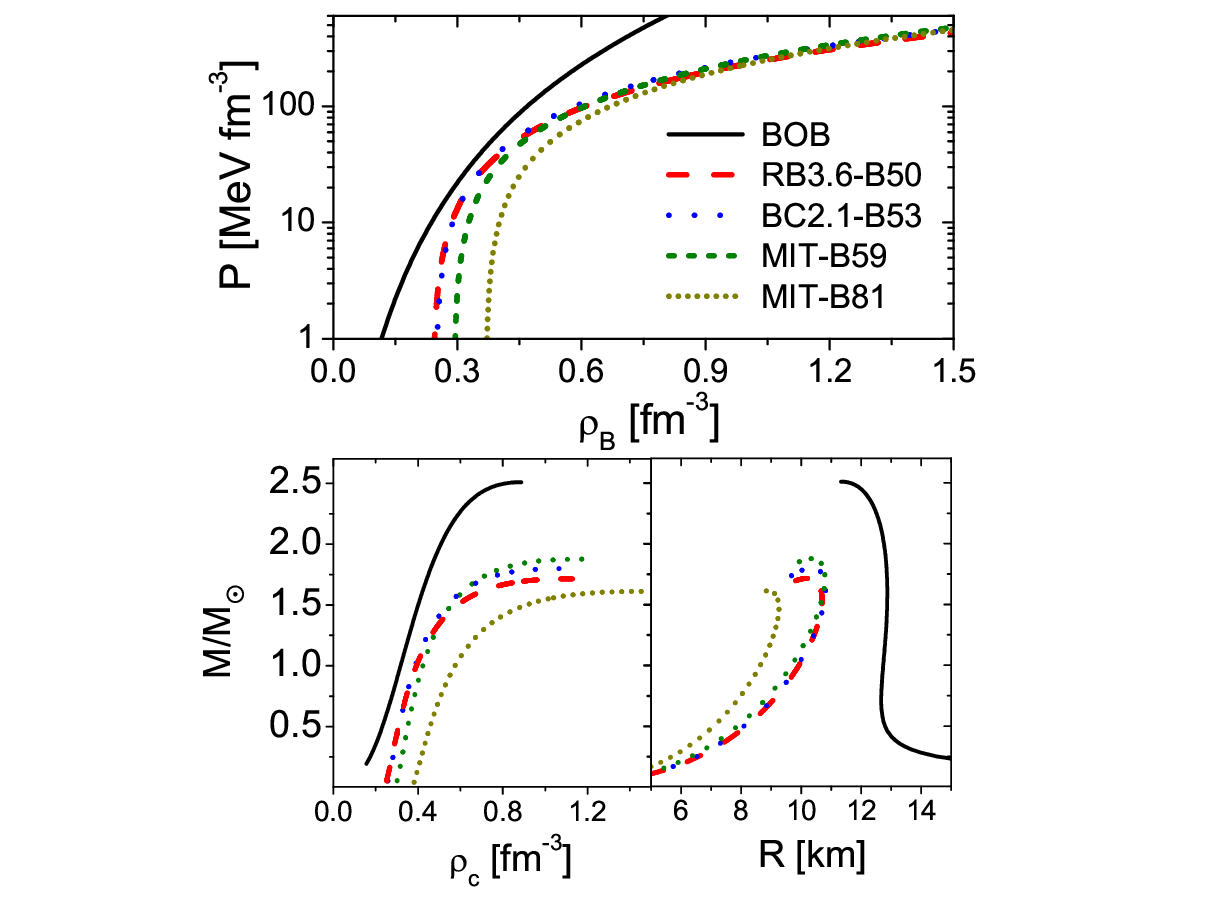}}
\vspace{-2mm}
\caption{(Color online)
Upper plot:
Pressure vs.~baryon number density of SQS matter
obtained with different EOSs.
Lower plots:
The corresponding gravitational mass -- central density and mass -- radius
relations of SQSs
($\ms=2\times10^{33}\text{g}$).
%and $\rho_0=0.17\;\text{fm}^{-3}$).
%The markers indicate the positions of the maxima.
See text for details.
}
\label{f:eos}
%\mycom{Space between var's and units.}
\end{figure}

The relevant EOS for SQSs is the one of beta-stable and charge-neutral SQM,
comprising a small electron fraction due to the finite strange quark mass,
and we present that EOS in the form $P(\rho_B)$ in Fig.~\ref{f:eos}(top)
for both the RB and BC models.
In each case, we choose three sets of parameters ($\al,\bds$),
corresponding to the three extreme points
of the allowed regions in Fig.~\ref{f:ab}.
Note that two of these points (MIT limit) coincide for RB and BC model.
For comparison, the nuclear BHF EOS is also shown.
Different from nuclear matter,
SQM approaches nonzero densities of about 0.25--0.4$\,\text{fm}^{-3}$
at zero pressure.
These are the surface densities of the corresponding SQSs,
which decrease with decreasing $\bds$,
i.e., increasing stability of QM.
At higher densities,
the pressure (and energy density) of SQM is much lower than that
of nuclear matter.

%...................................................................

\begin{figure}[t]%..............................................................
%\vspace{10mm}
\centerline{\includegraphics[bb=1 90 780 1070,scale=0.3]{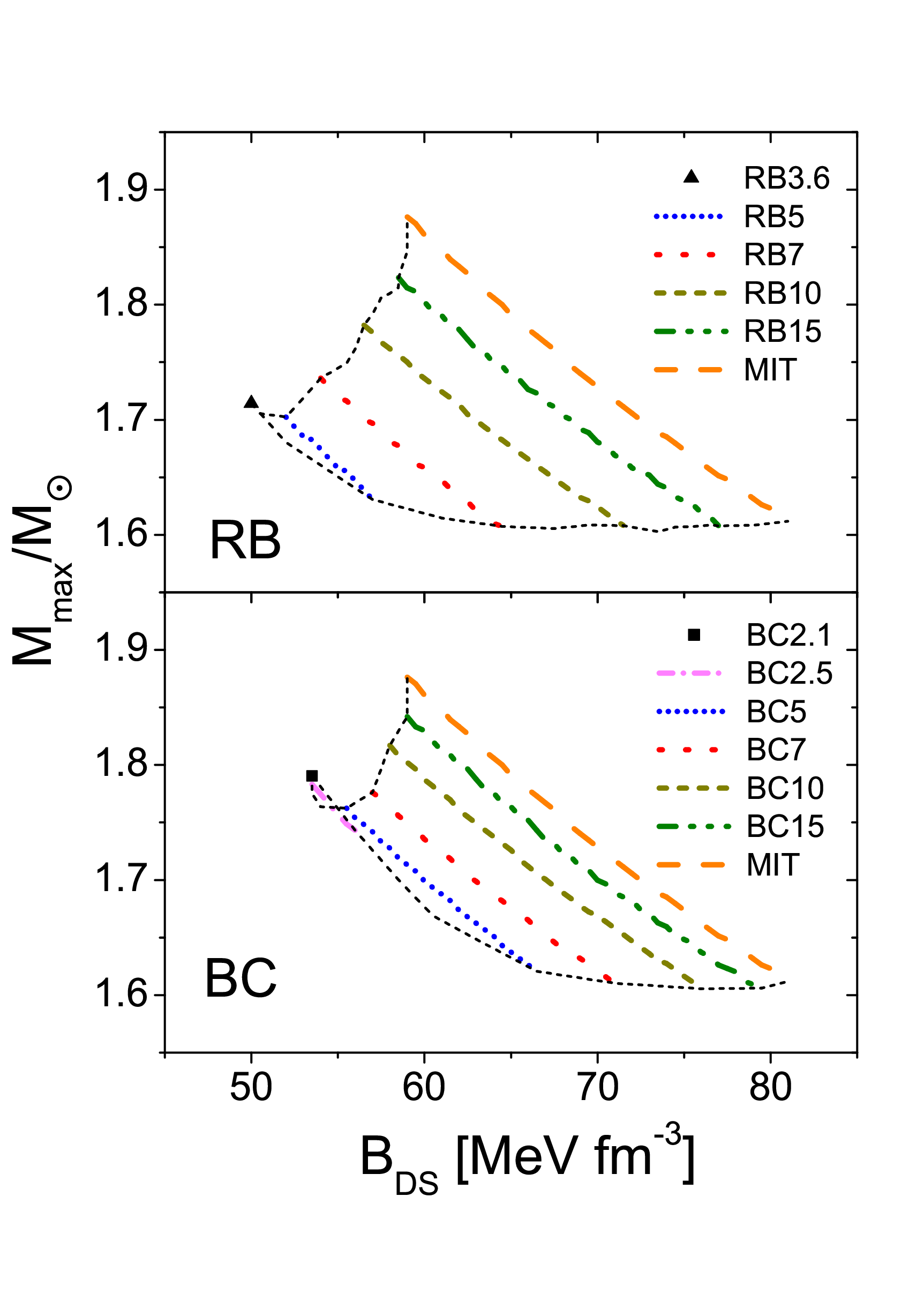}}
%\vspace{-7mm}
\caption{(Color online)
Maximum SQS mass
($\ms=2\times10^{33}\text{g}$)
in the DSM with RB (top panel)
and BC (bottom panel) approximation,
dependent on the value of $\bds$ for various fixed values of $\al$
within the constrained regions in Fig.~\ref{f:ab}.}
\label{f:Bmmax}
\end{figure} %..................................................................

With these EOSs we calculate the structure of bare SQSs.
We treat a SQS as a spherically symmetric distribution of
mass in hydrostatic equilibrium and
obtain the stellar radius $R$ and the gravitational mass $M$
by the standard procedure of solving the TOV equations \cite{shapiro}.
We assume SQSs without crust,
whereas for NSs we employ the BOB high-density BHF EOS joined with
the ones by Negele and Vautherin \cite{nv} in the medium-density regime,
and by Feynman-Metropolis-Teller \cite{fey} and
Baym-Pethick-Sutherland \cite{bps} for the outer crust.

The results are shown in the lower panels of Fig.~\ref{f:eos}.
The left (right) panel shows the star's gravitational mass dependent on
the central baryon density (radius).
Two of the quark model curves coincide with the MIT model,
and in fact the configuration with minimum $\bds$
provides the largest maximum mass and radius of all EOSs.
Increasing the bag constant or maintaining interacting quarks
with finite $\al^{-1}$,
reduces effectively the binding of QM and
thus the values of the above physical quantities.
In the lower right panel of Fig.~\ref{f:eos},
we find that typical radii of SQSs heavier than one solar mass are 10--11 km,
a bit smaller than hadronic NS radii 11--13 km,
but close to a recent analysis on observed compact stars \cite{Ozel15}.
Compared to other quark models \cite{Fraga01,Tian12},
we obtain a much smaller range of possible radii and maximum masses of SQSs.
This is due to the lower bound on the parameter $\bds$
from the constraint Eq.~(\ref{Eq:sn}).
For lower values of $\bds$ larger SQS masses can be reached,
but the stability condition of ordinary nuclear matter is violated.

The single most relevant number characterizing a given NS model
is probably the maximum mass,
in view of the recent observation of heavy NSs \cite{heavy,Fonseca16,heavy2}.
In order to address this issue,
we show in Fig.~\ref{f:Bmmax} the maximum mass of SQSs
for the allowed values of parameters $(\al,\bds)$,
as determined in Fig.~\ref{f:ab}.
In line with the repulsive character of the interaction in the DSM,
the maximum mass increases with increasing binding of QM, i.e.,
with decreasing value of $\bds$ for fixed $\al$,
or with increasing value of $\al$ for fixed $\bds$.

The globally largest value,
$M_\text{max}\approx1.9\,\ms$,
is thus found in the MIT limit ($\al^{-1}=0$)
with the smallest possible bag constant $B_\text{min}=59$.
This value is smaller than the currently observed largest mass of
compact stars \cite{heavy,heavy2}.
Regarding finally the dependence on the remaining parameter $m_s$,
it was shown in \cite{Farhi84,Torres13}
that in the MIT limit a decreasing value of $m_s$ produces
a larger permitted parameter space of the bag constant and
an increase of $M_\text{max}$; however,
even for a vanishing strange quark mass $M_\text{max}<1.95\,\ms$.

In comparison with SQSs,
the maximum mass of hybrid NSs can be much larger
at larger $\bds$ and/or smaller $\al$ \cite{Chen2011,Chen12,Chen15a,Chen15b}.
%The maximum mass increases with increasing $\bds$ or decreasing $\al$.
Due to the fact that a very stiff nuclear EOS (BOB)
was used for their construction,
in this case large maximum masses of about $2.5\,\ms$ can be reached,
which correspond to configurations where QM is only
present in the very core of the star.
However, this result cannot resolve the `hyperon puzzle' \cite{mmy,tom,Rijken16},
which would impede the existence of heavy hadronic or hybrid NSs,
where QM only appears at very large density.
We leave detailed discussions of different realizations of hybrid NSs
and their behavior in the whole parameter space to future work.

%For comparion we show in Fig.~???
%\mycom{new figure or 4 panels in Fig.5 ?}
%the maximum masses of hybrid NSs \cite{Chen2011,Chen2015}
%subject to the relevant parameter constraints of Fig.~\ref{f:ab}???.
%Now we give a short comparison between SQS and hybrid NSs.
%comparing with SQS, hybrid NSs are different
%As discussed above, absolutely stable SQM and SQSs can only exist
%in the grey parameter space of Fig.~\ref{f:ab}, while nuclear matter
%can transit to QM at certain density above the line of $B_{max}$.
%If the phase transition density is not too large,
%QM can present in the core of NSs.
%Such hybrids NS were studied in our previous work.
%However, if one exclude strange quark by hand,
%there can also exist a phase transition of nuclear matter to 2QM,
%even in the parameter space relevant to constraints of Fig.~\ref{f:ab}.
%In such cases, there could exist hybrid NSs with 2QM in the core of NSs.

%===============================================================================
\subsection{Conversion of neutron stars into strange quark stars}
\label{s:coll}

\begin{figure}[t]%..............................................................
\vspace{-11mm}
\centerline{\includegraphics[scale=0.42]{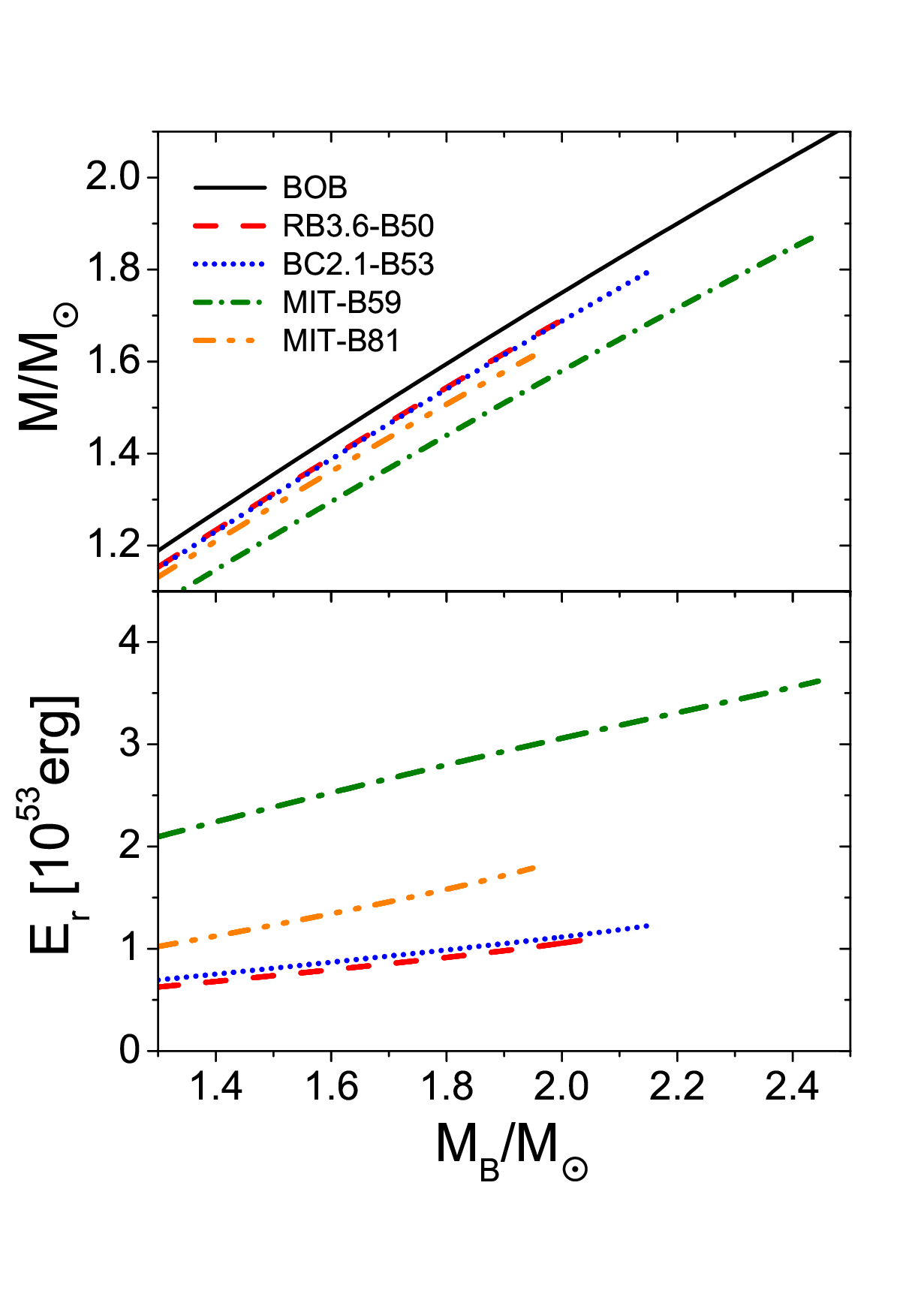}}
\vspace{-12mm}
\caption{(Color online)
Upper panel: Gravitational mass vs.~baryonic mass of compact stars
for different EOSs.
Lower panel: Energy release vs.~baryonic mass
in the conversion of NSs into SQSs with baryon number conservation.}
%($\ms=2\times10^{33}\text{g}$)}
\label{f:enre}
\end{figure} %..................................................................

As a last application,
we consider the formation of a SQS from a pure metastable NS,
which might occur once and if an initial seed of SQM has been formed
in one of several hypothetical ways \cite{Alcock86,seed,Iida98,Logoteta12},
for example in a two-step phase transition of nuclear matter to 2QM to SQM
\cite{Dai92,Niebergal10},
or in the combustion of hot nuclear matter \cite{burn,Pagliara13,Drago15,Ouyed15,Furusawa15a,Furusawa15b}.

This transition is accompanied by a huge energy release,
which could be associated with the long gamma-ray bursts
\cite{Cheng96,Bombaci00,Berezhiani03}
or the two-neutrino-burst scenario supernovas \cite{burst}.
Here we simply assume that the total baryon number is conserved
during the transition;
then the energy release can be obtained from the mass difference
between the pure NS and the SQS with the same total baryon number.

In Fig.~\ref{f:enre} we show the relation between gravitational mass $M$
and baryonic mass $M_B$ of pure NSs and SQSs (upper panel)
and the energy release $E_r$ in the conversion of NSs into SQSs
with baryon number conservation (lower panel).
The energy release depends on the DSM parameters and
for a typical NS with $M=1.4\,\ms$
can be %up to $0.12\,\ms$, i.e.,
(0.8--2.5)$\times 10^{53}\,\text{erg}$,
the maximum value of which is obtained for the MIT case ($\al^{-1}=0$)
with minimum value of the bag constant, i.e.,
for the strongest bound SQS,
which also allows the largest maximum mass of SQSs, see Fig.~\ref{f:eos}.

These results are quantitatively similar to others in the literature, e.g.,
a range of (1--4)$\times 10^{53}\,\text{erg}$ was obtained
in Ref.~\cite{Bombaci00}.
As the NS mass increases,
the energy release in the conversion also increases,
up to $3.6\times 10^{53}\,\text{erg}$
for a NS with $M=2.1\,\ms$ and the SQS obtained from the MIT-B59 case.
For NSs with even larger masses, $M\approx\;$(2.1--2.5)$\,\ms$,
there are no SQSs corresponding to the same baryon number.
Therefore, it is impossible to convert such heavy NSs into SQSs
with baryon number conservation,
and they can only be converted into SQSs with a big loss of baryons,
or into black holes.

%===============================================================================
\section{Conclusions}
\label{s:end}

We have investigated SQM and SQSs in our DSM for QM.
For the hypothesis of SQM to be valid, i.e.,
SQM being stable against nuclear matter,
while the latter is stable against 2QM,
we obtained the allowed parameter space of $\al$ and $\bds$
in our DSM with RB or BC vertex.
We found that SQM exists only for
fairly low values of the bag parameter $\bds$
and furthermore a sufficiently strong damping of the
in-medium effective interaction,
expressed by a lower limit on the parameter $\al$.
The strongest bound configurations
correspond to the MIT bag-model limit, i.e.,
a vanishing in-medium effective interaction,
with a minimal value of the bag constant.

This reinforces the idea that the full DSM
does not provide a sufficiently attractive in-medium quark-quark
interaction in order to create SQM.
In fact that interaction turns out strongly repulsive in the DSM.
This is in agreement with the NJL model, which does not
allow SQS either,
but in contrast to bag-model-type calculations
(eventually including perturbative QCD corrections),
which feature attractive interaction and strongly bound SQM
with suitable parameter choices.

Outside the SQM parameter limits,
hybrid two- or three-flavor NSs may exist in a much larger domain
of the DSM parameter space,
because in this case the upper limit on $\bds$ is established
by a hadron-quark phase transition at high density in the NS core.
Metastable two-flavor hybrid NSs may even coexist with SQSs.
We will study these aspects in future work.

Then, in the allowed parameter space,
we calculated the EOS of SQM and the corresponding structure of SQSs.
We found that the maximum mass of SQSs is about $1.9\,\ms$
and typical radii are 9--11 km,
while maximum masses of ordinary (hybrid) NSs are
$2.5\,\ms$ and 11--13 km,
depending on the hadronic EOS, though.
We finally discussed the formation of SQSs due to the conversion from NSs
and obtained energy releases
as large as $3.6 \times 10^{53}\,\text{erg}$.

For the future it will be important to establish direct estimates of
the DSM model parameters in a more fundamental way from QCD,
and in this way to clarify the qualitative differences between
the different quark models mentioned above.

\section*{Acknowledgments}

We acknowledge
useful comments from D.~Blaschke and
financial support from NSFC (11305144,11303023),
Central Universities (CUGL 140609) in China.
Partial support comes from ``NewCompStar," COST Action MP1304.

%%%%%%%%%%%%%%%%%%%%%%%%%%%%%%%%%%%%%%%%%%%%%%%%%%%%%%%%%%%%%%%%%%%%%%%%%%%%%%%%

\end{document}